\documentclass[conference]{IEEEtran}
\IEEEoverridecommandlockouts
\usepackage{cite}
\usepackage{amsmath,amssymb,amsfonts}
\usepackage{graphicx}
\usepackage{textcomp}
\usepackage{xcolor}
\usepackage{subcaption}
\usepackage{tikz}
\usepackage{pgfplots}
\usepackage{booktabs}
\usepackage{algorithm}
\usepackage{makecell}
\usepackage{multirow}
\usepackage[noend]{algpseudocode}
\usepackage[a4paper, total={184mm,239mm}]{geometry}
\usepackage{url}
\def\BibTeX{{\rm B\kern-.05em{\sc i\kern-.025em b}\kern-.08em
    T\kern-.1667em\lower.7ex\hbox{E}\kern-.125emX}}

\usepackage{pifont}
\newcommand{\cmark}{\ding{51}}%
\newcommand{\xmark}{\ding{55}}%

\definecolor{jcred}{HTML}{e31a1c}
\definecolor{jcgreen}{HTML}{33a02c}
\definecolor{jcblue}{HTML}{1f78b4}
\definecolor{jcorange}{HTML}{ff7f00}
\definecolor{jcpurple}{HTML}{6a3d9a}
\definecolor{jclightred}{HTML}{fb8072}
\definecolor{jclightgreen}{HTML}{b3de69}
\definecolor{jclightblue}{HTML}{80b1d3}
\definecolor{jclightorange}{HTML}{fdb462}
\definecolor{jclightpurple}{HTML}{bebada}
\definecolor{jcredl}{HTML}{fb8072}
\definecolor{jcgreenl}{HTML}{b3de69}
\definecolor{jcbluel}{HTML}{80b1d3}
\definecolor{jcorangel}{HTML}{fdb462}
\definecolor{jcpurplel}{HTML}{bebada}

\begin{document}

\title{HASS: Hardware-Aware Sparsity Search for Dataflow DNN Accelerator}
\author{\IEEEauthorblockN{Zhewen Yu\IEEEauthorrefmark{1},  Sudarshan Sreeram\IEEEauthorrefmark{1},  Krish Agrawal\IEEEauthorrefmark{1}, Junyi Wu\IEEEauthorrefmark{1}, Alexander Montgomerie-Corcoran\IEEEauthorrefmark{1}, \\ Cheng Zhang\IEEEauthorrefmark{1},  Jianyi Cheng\IEEEauthorrefmark{2}, Christos-Savvas Bouganis\IEEEauthorrefmark{1}, Yiren Zhao\IEEEauthorrefmark{1}}
\IEEEauthorblockA{
    \IEEEauthorrefmark{1}Imperial College London, UK, 
    \IEEEauthorrefmark{2}University of Cambridge, UK \\
    \{zhewen.yu18, sudarshan.sreeram19, krish.agrawal20, junyi.wu21, alexander.montgomerie-corcoran15,\\ cheng.zhang122, christos-savvas.bouganis, a.zhao\}@imperial.ac.uk, jianyi.cheng@cl.cam.ac.uk}
}

\maketitle

\begin{abstract}
Deep Neural Networks (DNNs) excel in learning hierarchical representations from raw data, such as images, audio, and text. To compute these DNN models with high performance and energy efficiency, these models are usually deployed onto customized hardware accelerators. Among various accelerator designs, dataflow architecture has shown promising performance due to its layer-pipelined structure and its scalability in data parallelism.

Exploiting weights and activations sparsity can further enhance memory storage and computation efficiency. However, existing approaches focus on exploiting sparsity in non-dataflow accelerators, which cannot be applied onto dataflow accelerators because of the large hardware design space introduced. As such, this could miss opportunities to find an optimal combination of sparsity features and hardware designs. 

In this paper, we propose a novel approach to exploit unstructured weights and activations sparsity for dataflow accelerators, using software and hardware co-optimization. We propose a Hardware-Aware Sparsity Search (HASS) to systematically determine an efficient sparsity solution for dataflow accelerators. Over a set of models, we achieve an efficiency improvement ranging from 1.3$\times$ to 4.2$\times$ compared to existing sparse designs, which are either non-dataflow or non-hardware-aware. Particularly, the throughput of MobileNetV3 can be optimized to 4895 images per second. HASS is open-source: \url{https://github.com/Yu-Zhewen/HASS}
\end{abstract}

\section{Introduction}
Deep Neural Networks (DNNs) models are designed to extract relevant features from raw data, such as images, audio, and text. To improve the performance and energy efficiency when computing these models, the computation is often mapped onto hardware accelerators. Among various accelerator architectures, dataflow accelerators have shown significant performance benefits because of their deeply pipelined computation between layers~\cite{venieris_fpgaconvnet_2018} and scalable data parallelism across devices~\cite{9499943}.

In hardware accelerator designs, sparsity has been a popular topic, such that unnecessary computation with zeros can be avoided for better efficiency. There are two types of sparsity:

\begin{itemize}
    \item \textbf{Weight Sparsity}: focusing on the zeros in the weights of a model, whose positions are often available at compile time so that they can be optimized statically.
    
    \item \textbf{Activation Sparsity}: focusing on the zeros inside the intermediate activation data between layers. The positions of these zeros are only known at run-time, as they depend on the network input.
\end{itemize}

With the appropriate hardware support, increased sparsity can lead to fewer computations. To maximize the sparsity, pruning is a process that simplifies a model by setting certain weights and activations to zeros, at minimal accuracy loss based on a set of criteria \cite{jain2020domain}. 

\begin{figure}
    \centering
    \begin{tikzpicture}[thick, scale=0.95, every node/.style={scale=0.82}]
\pgfplotsset{
    tick label style={font=\large},
    label style={font=\large}, 
    compat=1.18
}
    \begin{axis}[
        xlabel={Operations Density (images/cycle/DSP)},
        ylabel={Top1 Accuracy (\%)},
        height=65mm,
        width=0.48\textwidth,
    ]
    
    \addplot[mark=diamond, mark size=3, black!30!green, dashed, mark options={solid}] plot coordinates {
        (3986/5346/250000000, 71.786)
        (3986/5152/250000000, 71.748)
        (4495/5261/250000000, 71.448)
        (4610/5220/250000000, 71.120)
    };
    \node[above, text=black!30!green] at (3986/5346/250000000, 71.786) {HASS};

    \addplot[draw=none, mark=diamond, mark size=3, black!30!green, mark options={solid}]plot coordinates {
        (3986/5098/250000000, 71.404)
        (3986/5336/250000000, 71.654)
        (3986/5072/250000000, 71.412)
        (3986/4938/250000000, 71.182)
        (3986/4815/250000000, 70.944)

        (1035/1783/250000000, 71.182)
        (1015/1766/250000000, 71.412)
        (963/1613/250000000, 71.448)
        (1027/1729/250000000, 71.406)        (1035/1783/250000000, 71.182)
        (1035/1742/250000000, 70.944)

        (1135/1783/250000000, 71.282)
        (1215/1766/250000000, 71.382)
        (1063/1613/250000000, 71.048)
        (1097/1729/250000000, 71.406)        (1235/1783/250000000, 71.182)
        (1135/1742/250000000, 70.944)

    };
    
    \addplot[draw=none, mark=star, mark size=3, orange, mark options={solid}]plot coordinates {
        (1660/3596/250000000, 71.786)
    };
    \node[below, text=orange] at (1660/3596/250000000, 71.786) {PASS \cite{montgomerie2023pass}};

    \addplot[draw=none, mark=+, mark size=3, blue, mark options={solid}]plot coordinates {
        (4539/5928/390000000, 71.90)
    };
    \node[above, text=blue] at (4539/5928/390000000, 71.90) {HPIPE \cite{hall2020hpipe}};
    
    \addplot[draw=none, mark=o, mark size=3, red, mark options={solid}]plot coordinates {
        (302.3/2160/150000000, 70.80)
    };
    \node[above, text=red] at (302.3/2160/150000000, 70.80) {\cite{liu2023efficient}};
    \end{axis}
\end{tikzpicture}
    \caption{HASS explores the optimal trade-off between classification accuracy and operation density. We compare HASS with other existing sparse MobileNetV2 implementations. }
    \label{fig:performance_density}
\end{figure}
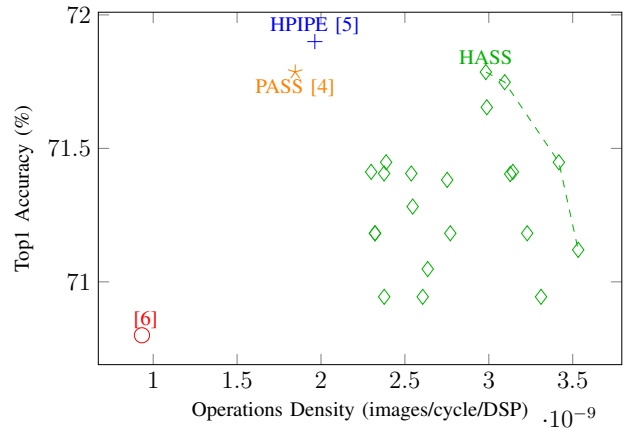

\begin{table}
    \centering
    \caption{Our work is the first attempt to exploring hardware-aware unstructured pruning for dataflow accelerators. Our approach systematically exploits layer-wise sparsity in both weights and activations coupled with hardware resource-constrained analysis.}
    \label{tab:high-level-comparison}
    \resizebox{\columnwidth}{!}{%
    \begin{tabular}{lccccccccc}
    \toprule 
       Approaches & \cite{lu2018spwa} & \cite{lu2019efficient} & \cite{zhu2020efficient} & \cite{hall2020hpipe} &\cite{liu2023efficient} & \cite{kong2022spvit} & \cite{qu2023speed} &  \cite{montgomerie2023pass}&  Ours \\
    \midrule
        Publication Year & 2018 & 2019 & 2020 & 2020 & 2022 & 2023  & 2023 &  2023 & 2024\\
        Dataflow architecture & \color{jcred}{\xmark} & \color{jcred}{\xmark} & \color{jcred}{\xmark} & \color{jcgreen}{\cmark} & \color{jcred}{\xmark} & \color{jcred}{\xmark}  & \color{jcred}{\xmark}  & \color{jcgreen}{\cmark} & \color{jcgreen}{\cmark}  \\
        Weight sparsity & \color{jcgreen}{\cmark} & \color{jcgreen}{\cmark} & \color{jcgreen}{\cmark} & \color{jcgreen}{\cmark} & \color{jcgreen}{\cmark} & \color{jcgreen}{\cmark}  & \color{jcgreen}{\cmark} & \color{jcred}{\xmark} & \color{jcgreen}{\cmark} \\
        Activation sparsity & \color{jcred}{\xmark} & \color{jcred}{\xmark} & \color{jcgreen}{\cmark} & \color{jcred}{\xmark} & \color{jcred}{\xmark}  &  \color{jcgreen}{\cmark}  & \color{jcgreen}{\cmark} & \color{jcgreen}{\cmark} & \color{jcgreen}{\cmark} \\
        Hardware-aware & \color{jcred}{\xmark} & \color{jcred}{\xmark} & \color{jcred}{\xmark} & \color{jcred}{\xmark} & \color{jcgreen}{\cmark} & \color{jcred}{\xmark}  & \color{jcgreen}{\cmark} & \color{jcred}{\xmark} & \color{jcgreen}{\cmark} \\
    \bottomrule
    \end{tabular}
    }
\end{table}
\begin{figure*}
    \centering
    \begin{subfigure}[b]{\textwidth}
    \centering
    \includegraphics[scale=0.55]{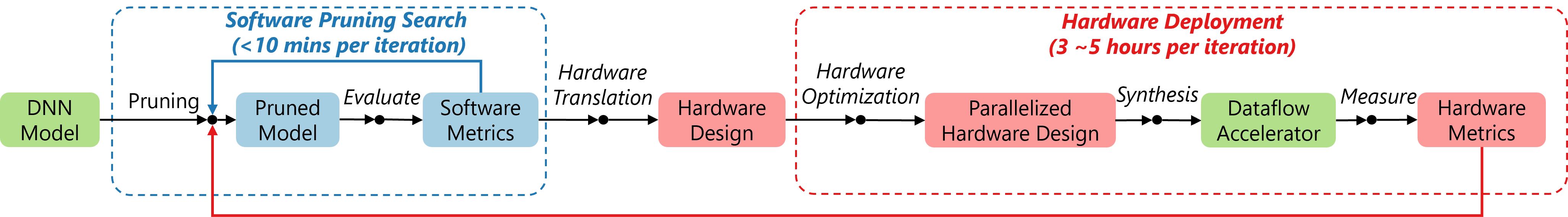}
    \caption{Traditional pruning flow for dataflow accelerator design focuses on software metrics only, where a pruned model is treated `read-only' in the hardware end.}
    \label{fig:base_flow}
    \end{subfigure}
    \begin{subfigure}[b]{\textwidth}
    \centering
    \includegraphics[scale=0.55]{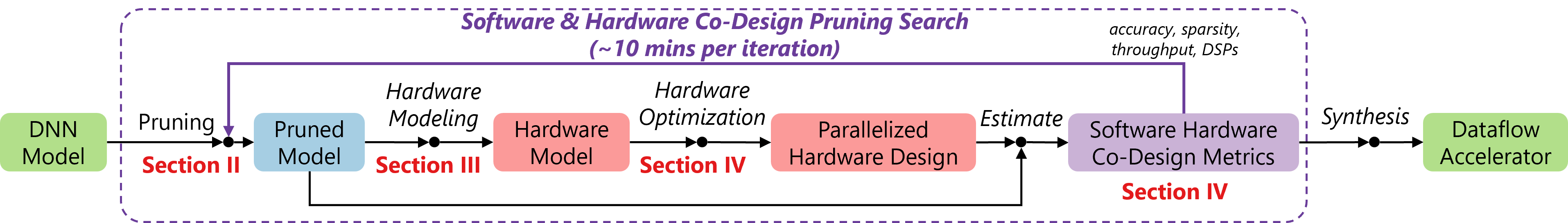}
    \caption{Our flow integrates hardware metrics into the pruning search process. We will explain the details of each step in the highlighted sections.}
    \label{fig:our_flow}
    \end{subfigure}
    \caption{An overview of our approach. When designing a dataflow accelerator, traditional pruning steps are separate from the hardware synthesis steps, which restricts software and hardware co-optimization. Our flow explores both software and hardware optimization concurrently, opening up opportunities to find the optimal design.}
    \label{fig:overview}
\end{figure*}

Traditional pruning approaches only consider software metrics, such as the overall sparsity in the network, without directly considering the actual impact on hardware performance. Recently, there has been interest in developing hardware-aware, co-design pruning approaches for accelerator designs. These approaches consider hardware performance, such as throughput and energy efficiency, into the pruning criteria. These approaches realize co-optimization of both accuracy and hardware performance.

In non-dataflow hardware accelerators, sparsity is exploited to lift the performance bottleneck in the off-chip memory bandwidth 
\cite{jain2023modular}. To efficiently represent sparse data, encoding techniques are often used to reduce both the memory footprints and the required off-chip memory bandwidth\cite{parashar2017scnn}. Pruning can be further tailored to optimize individual sparse computation for efficient data access and processing \cite{dave2021hardware}. 

However, these approaches cannot be applied to the design of dataflow accelerators for two reasons. First, the performance bottleneck for dataflow accelerators is often the computation resources rather than the off-chip memory bandwidth, because most weights and activations reside on-chip. More importantly, in a sparse dataflow accelerator, the overall pipeline performance does not scale proportionally with the total non-zero operations. For example, increasing the sparsity of a non-slowest layer will not change the pipeline performance at all, unless it allows for the reallocation of resources to alleviate the pipeline bottleneck \cite{montgomerie2023pass}.


As such, exploiting the sparsity in a dataflow architecture poses unique challenges in hardware designs, including scheduling and resource allocation within a given hardware resource budget. Existing approaches on sparse dataflow architecture separate the pruning steps and hardware optimization steps, as shown in Fig.~\ref{fig:base_flow}. In this paper, we propose a systematic approach to explore unstructured pruning and hardware optimization in a co-design form, as shown in Fig.~\ref{fig:our_flow}. To the best of our knowledge, our approach is the first attempt at hardware-aware pruning for dataflow accelerators, as illustrated in Table~\ref{tab:compare}. Our main contributions are as follows.
\begin{itemize}
    \item The first dataflow accelerator design that exploits both weight sparsity and activation sparsity coupled with layer-pipelined execution.
    \item A hardware-aware, unstructured model pruning algorithm that considers both software pruning metrics (accuracy, sparsity ratio) and hardware performance (throughput, resource estimations) for systematic optimization.
    \item Over a set of DNN models, we show that our approach leads to an improved efficiency ranging from 1.3$\times$ to 4.2$\times$ compared to existing sparse designs.
\end{itemize}

\section{Related Work}
\label{sec:related_work}

In the existing studies, the sparse DNN-FPGA accelerators are often implemented as a single sparse-sparse matrix multiplication engine, shared by DNN layers in a time-multiplexed manner. As such, the main challenge is irregular memory access patterns within and between layers. As such, sparse data are often encoded to save memory space and simplify the scheduling of non-zero values to processing units \cite{parashar2017scnn}. Lu \textit{et al.} \cite{lu2019efficient} explored the weight sparsity only. As the positions of zeros are known at compile-time, a look-up table is built to match the indices of sparse weights and dense activations. To exploit the dynamic, data-dependent activation sparsity, Zhu \textit{et al.} \cite{zhu2020efficient} used the activation value to control the clock gating of processing units, which leads to energy saving but the achieved throughput remains the same as the dense computation. As irregular sparsity pattern leads to inefficient hardware, Kong \textit{et al.} \cite{kong2022spvit} proposed the latency-aware pruning that jointly optimizes the accuracy and latency. Similarly, Qu \textit{et al.} \cite{qu2023speed} modelled energy consumption and processing cycles, and considered them together in the loss function, so that the accelerator architecture can be considered.

Instead of scheduling layers in a time-multiplexed manner, HPIPE \cite{hall2020hpipe} and PASS \cite{montgomerie2023pass} are two works that look in a different direction, building the dataflow architecture that has multiple pipelined sparse matrix multiplication engines. To solve the resource allocation and throughput balancing between multiple sparse engines, they estimated the distribution of sparse data in PyTorch and used that estimation to guide the process of design space exploration. However, HPIPE only exploits the weight sparsity, while PASS only exploits activation sparsity, and neither of them has considered the hardware-aware co-design in their approaches.

\section{Unstructured Pruning Algorithm}
\label{sec:pruning_algorithm}

\begin{figure*}[t]
    \centering
    \includegraphics[width=0.9\textwidth]{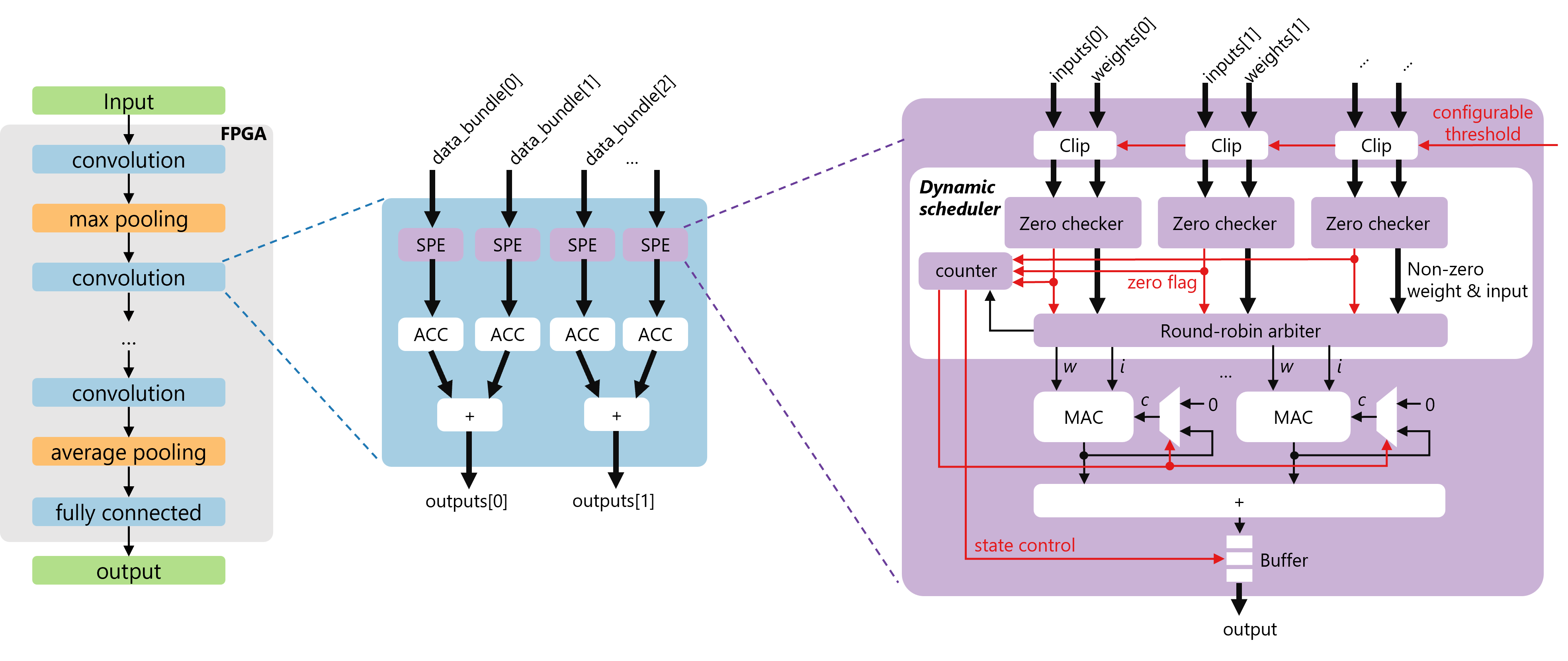}
    \caption{Architecture of the sparse dataflow accelerator. The computation of layers is pipelined where inside each layer, there are multiple Sparse Vector Dot Product Engines (SPE) operating in parallel. Inside the SPE, clipping modules zero out any weight or activation that falls below a configurable threshold. Zero-filtering then detects these zeros, while the remaining non-zeros are dispatched to the MACs (implemented with DSPs) via an arbiter. We also use a dedicated counter to track the number of skipped zeros and manage accumulation result emissions.}
    \label{fig:sparse_engine}
\end{figure*}

In this paper, we choose to apply one-shot pruning without fine-tuning, where a significant portion of the network is pruned in a single step, to reduce the deployment effort. This is in contrast to iterative pruning approaches, where pruning is performed gradually over multiple iterations, accompanied by fine-tuning stages \cite{tan2020dropnet}. 

In terms of the pruning criterion, we use the magnitude-based evaluation (L1 norm), where weights and activations having smaller magnitudes below the threshold will be forced to zeros. The pruning threshold can either be uniform or unique across layers. Adopting a uniform threshold is straightforward to explore the trade-off between accuracy and sparsity. However, setting unique thresholds instead better preserve the network accuracy, as many researchers have demonstrated the diversity of statistics per layer \cite{banner2019post, dave2021hardware}. 

Therefore, for each individual layer in the network, our pruning algorithm requires the identification of unique thresholds $\tau_w$ and $\tau_a$ for weight and activation pruning respectively, and the introduced sparsity is denoted as $S_w$ and $S_a$, with a range between 0 and 1, indicating the fraction of zeros in the data. As introduced before, once the pruning threshold is given, the weight sparsity $S_w$ is a fixed value since the positions of zeros are already available at compile time. However, the activation sparsity $S_a$ depends on the network's input, making it dynamic at run-time.

\section{Sparse Dataflow Architecture}

Traditional sparse architectures use two approaches to address the challenges in memory storage and computational efficiency. For efficient memory storage, they statically encode weights, trading off between efficient memory storage and runtime overhead caused by decoding \cite{parashar2017scnn}. For efficient computation, they skip zeros during computation, but leaving the corresponding hardware operator idle \cite{zhu2020efficient}. To fully utilize the available computation resources, we statically analyze the run-time sparsity and pre-fetch data in a buffer to keep the hardware operators busy at each cycle. This leads to better performance than the zero-skipping approach. 

Fig.~\ref{fig:sparse_engine} illustrates our target sparse dataflow accelerator architecture. On the left, the model is presented in a dataflow graph, where each node represents a dataflow component in hardware and each edge represents their data interface between every two nodes. The blue nodes are often resource intensive, such as DSP blocks on an FPGA, and can be significantly optimized by sparse computation.
In each blue node, data parallelism is applied to improve performance. For example, the middle of the figure represents a hardware convolutional layer. The hardware optimization steps consider two metrics: 1) spatial data parallelism by duplicating hardware processing elements, shown as multiple Sparse Vector Dot Product Engine (SPE) in the figure, and 2) time-wise folding by iterating using the same SPE and accumulating the result, shown as ACC and ``+'' in the figure. Detailed exploration on these metrics are explained in Section~\ref{subsec:thr}.

The hardware implementation of sparse computation is in SPE, shown in the right side of Fig.~\ref{fig:sparse_engine}. The engine takes a set of inputs from the preceding layer and weights from the on-chip memory, which are passed through the clip modules which would zero out any value below the configurable threshold. Afterwards, zero values are directly forwarded to the counter on the left to keep track of the iteration count, while non-zero pairs are buffered in a round-robin arbiter for computation. The arbiter dispatches multiple pairs to the available Multiply–Accumulate (MAC) units concurrently such that they are busy in each clock cycle. Once the counter reaches full, the output data will be released. 

Note that the accumulation of the vector product can take place both inside each SPE and also between multiple SPEs. This strategy can constrain the fan-in and fan-out of the arbiter, to reduce the area overhead and the degradation of clock frequency.

The number of the MACs in each SPE, denoted as $N$, is statically determined by estimating the overall sparsity of the input activations and weights on a calibration dataset. Consider the scenario where a SPE takes $M$ input weight pairs. In traditional hardware architecture for dense computation, the computation would take $M/N$ cycles to complete. Each MAC unit accumulates for $M/N$ times before emitting partial sums to the $N$-input adder tree for the final output calculation. In sparse hardware architecture, let $\overline{S}$ be the average sparsity of the input activation and weight pair, where the computations can be skipped if any s any of them is equal to zero. In this case, the arbiter will schedule the computation to complete with $t$ cycles, where:
\begin{equation}
    t(\overline{S}) = \lceil \frac{(1-\overline{S}) \times M}{N} \rceil
\end{equation}
This is also known as initial interval of the SPE. Considering both weight and activation sparsity, our focus is on the probability of either weight or activation becoming zero. In our implementation, $N$ and $M$ are customized for each layer according to its $\overline{S}$.

The proposed sparse computation engine can be aggregated to enhance performance, both within a layer (intra-layer) and between layers (inter-layer). Within each convolutional layer, concurrent vector dot products can occur, allowing parallel computation across the input-channel ($I$) and output-filter ($O$) dimensions. We designate the levels of this parallelism as $i \in [1,I]$ and $o \in [1,O]$, respectively. In the case of unstructured pruning, the sparsity pattern is not uniform within a convolutional layer. Consequently, the processing rates of $i \times o$ SPEs are dynamic and may be imbalanced at run-time, potentially causing pipeline stalls. To mitigate this, we employ the following strategies:
\begin{itemize}
    \item \textbf{Balancing Strategy}: During compile-time, we estimate the weight and activation sparsity in each input-channel and output-filter. We then utilize simulated annealing to solve an allocation problem. This assigns the computation of $I$ input-channels and $O$ output-filters to $i \times o$ sparse computation engines, minimizing the difference in their processing rates.
    \item \textbf{Buffering Strategy}: Buffering is employed to absorb the instantaneous variance of dynamic processing rates. The selection of buffer size involves a trade-off between resource usage and throughput, and we determine the buffer size following a heuristic approach similar to \cite{montgomerie2023pass} which is based on the observation of moving window statistics.
\end{itemize}
In the sparse dataflow architecture, computation is pipelined on a layer-by-layer basis using FIFOs and handshake signals. In this architecture, activation data is not encoded despite its sparsity. This is because most intermediate data remains on-chip, and frequent encoding and decoding between layers would incur significant computational costs. 

\section{Hardware-aware Workflow}
\subsection{Accelerator Design Space Exploration}
\label{subsec:thr}

This section focuses on the Design Space Exploration (DSE) problem of the sparse DNN-FPGA accelerator, and it can be formalized using the following terms:
\begin{itemize}
    \item $L : \{l_0, l_1, l_2...\}$ denotes the layers in a network;
    \item $S \subseteq [0, 1)$ denotes the sparsity search space;
    \item $D : \{d_0, d_1, d_2...\}$ denotes all possible hardware design points of a layer;
    \item $g \subseteq L \times D \times S$ denotes a design point of the network;
\end{itemize}
Given a budget for the hardware resources $R$, we search for an efficient $g$ to maximize network throughput in a greedy form.

\subsubsection{Performance modeling}

Let $C_l$ denote the number of operations (including the zeros) in layer $l$. The throughput of the same layer is then:
\begin{equation}
    \theta(l, d, \overline{S})  = \frac{i \times o \times M}{C_l \times t(\overline{S})}
\end{equation}
Since $t$ is dynamic, depending on the average sparsity $\overline{S}$, so as the throughput of the layer. The sparse dataflow architecture is layer-wise pipelined, therefore, the network throughput $\theta$ is restricted by the slowest layer:
\begin{equation}
   \forall g \ldotp \theta \leq \min_{l \in L} \theta(l, d, \overline{S})
   \label{equ:thr}
\end{equation}
The aim of the DSE process is to determine an efficient $g$ with a large $\theta$. 

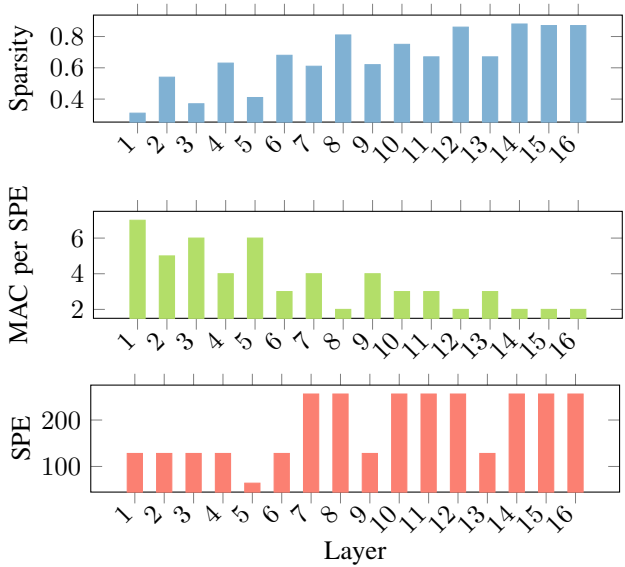
\begin{figure}[t]
    \centering
    \begin{tikzpicture}
        \begin{axis}[
            ybar,
            width=0.95\columnwidth,
            height=3cm,
            bar width=0.2cm,
            ylabel={Sparsity},
            y label style={at={(0.08,0.5)},anchor=south},
            xticklabel style={rotate=45,anchor=east},
            xtick=data,
            ]
            \addplot [jclightblue, draw=jclightblue, fill=jclightblue] coordinates {
                (1, 0.31) (2, 0.54) (3, 0.37) (4, 0.63) (5, 0.41) (6, 0.68) (7, 0.61) (8, 0.81)
                (9, 0.62) (10, 0.75) (11, 0.67) (12, 0.86) (13, 0.67) (14, 0.88) (15, 0.87) (16, 0.87)
            };
        \end{axis}
    \end{tikzpicture}
    \begin{tikzpicture}
        \begin{axis}[
            ybar,
            width=0.95\columnwidth,
            height=3cm,
            bar width=0.2cm,
            ylabel={MAC per SPE},
            y label style={at={(0.08,0.5)},anchor=south},
            xticklabel style={rotate=45,anchor=east},
            xtick=data,
            ]
            \addplot [jclightgreen, draw=jclightgreen, fill=jclightgreen] coordinates {
                (1, 7) (2, 5) (3, 6) (4, 4) (5, 6) (6, 3) (7, 4) (8, 2)
                (9, 4) (10, 3) (11, 3) (12, 2) (13, 3) (14, 2) (15, 2) (16, 2)
            };
        \end{axis}
    \end{tikzpicture}
    \begin{tikzpicture}
        \begin{axis}[
            ybar,
            width=0.95\columnwidth,
            height=3cm,
            bar width=0.2cm,
            xlabel={Layer},
            ylabel={SPE},
            y label style={at={(0.08,0.5)},anchor=south},
            xticklabel style={rotate=45,anchor=east},
            xtick=data,
            ]
            \addplot [jclightred, draw=jclightred, fill=jclightred] coordinates {
                (1, 128) (2, 128) (3, 128) (4, 128) (5, 64) (6, 128) (7, 256) (8, 256)
                (9, 128) (10, 256) (11, 256) (12, 256) (13,128) (14, 256) (15, 256) (16, 256)
            };
        \end{axis}
    \end{tikzpicture}
    \caption{DSE results of a specific sparse ResNet-18 workload with 16 $3\times3$ convolutional layers. The allocation of MAC per SPE mainly depends on the per-layer sparsity statistic. A higher sparsity leads to a smaller MAC per SPE. However, the increase of the layer index leads to an increasing number of convolutional filters, so as the number parallel SPEs to match the rate between layers.}
\end{figure}
.

\subsubsection{Resource-constrained rate balancing}

Apart from the slowest layer, the remaining layers in the pipeline may underperform, meaning that their actual throughput is significantly lower than the maximum achievable throughput using the allocated resources. These hardware resources could be unused at run-time, leading to inefficiency. To address this, we can reduce the parallelism of these layers without affecting overall throughput to enable more parallelism at the performance bottleneck. This is also known as rate balancing.

Let $\theta_r(l,d, \overline{S})$ represent the actual throughput of a layer instance $(l,d, \overline{S})$. After rate balancing, each layer in the balanced design $g'$ is configured with a parallelism level close to its actual throughput.
\begin{align}
 \theta_l' = \min \{\theta(l,d', \overline{S}) | \theta(l,d', \overline{S}) \geq \theta_r(l,d, \overline{S})\} \\
 \forall l \in L \wedge (l, d, \overline{S}) \in g \Rightarrow \theta(l, d', \overline{S}) = \theta_l' \wedge (l, d',\overline{S}) \in g' \label{eqn:balance}
\end{align}
Constraint~\eqref{eqn:balance} ensures that all layers in the design are computing efficiently in a pipeline, optimizing the overall resource utilization.

\subsubsection{Resource-constrained incrementing}

The DSE starts with the resource-minimal design, where the computation within each layer is fully sequential. At each iteration, the DSE increases the parallelism of the slowest layer by a small step followed by the rate balancing process described in step 2). This incremental process repeats until the resource budget $R$ satisfies. The resource utilization of each sparse computation engine is modeled on the basis of the regression model.

\subsubsection{Partitioning and reconfiguration}

In practice, the restriction caused by finite hardware resources could fail to map an entire DNN to a single dataflow accelerator device. In this work, we fold the dataflow pipeline at the block level and iteratively compute them on the available hardware resources using full reconfiguration on an FPGA device. This allows changing hardware architecture at run-time at the price of additional reconfiguration time. To reduce such overhead caused by the reconfiguration time, the data is processed in a large batch size \cite{venieris_fpgaconvnet_2018}. The decisions of where to split the partition and the number of partitions are given by a simulated annealing solver that trade-off the reconfiguration time and data parallelism gained.

\subsection{Multi-objective Search}
\begin{figure}[t]
    \centering
    \begin{tikzpicture}[thick,scale=0.95, every node/.style={scale=0.95}]
\pgfplotsset{every x tick label/.append style={font=\footnotesize}, compat=1.3}
\begin{semilogyaxis}[
    height=60mm,
    width=0.45\textwidth,
    xlabel={\footnotesize Number of Iteration Steps},
    ylabel={\footnotesize  images/cycle/DSP ($10^{-9}$) },
    xmin=1, xmax=100,
    ytick={0.8, 0.9},
    ymode=log, log ticks with fixed point, 
    legend cell align={left},
    legend style={at={(0.45,0.3)},anchor=north west, draw=none, fill=none, font=\footnotesize},
    ]

\addplot[jcblue, draw=jcblue, mark=*, line width=1pt, mark options={scale=0.7}] coordinates {
(1, 0.20089053846205265/250000000*1000000000)
(3, 0.214764832481035/250000000*1000000000)
(39, 0.2159211910020882/250000000*1000000000)
(48, 0.21608334955543024/250000000*1000000000)
(57, 0.22661955179402837/250000000*1000000000)
(96, 0.22661955179402837/250000000*1000000000)
};
\addlegendentry{Software metrics-only}

\addplot[jcgreen, draw=jcgreen, mark=*, line width=1pt, mark options={scale=0.7}] coordinates {
(1, 0.19853448303570642/250000000*1000000000)
(2, 0.20601927940513548/250000000*1000000000)
(10, 0.21489302670194604/250000000*1000000000)
(31, 0.22112638274371121/250000000*1000000000)
(34, 0.2214602856595072/250000000*1000000000)
(67, 0.22181710575474559/250000000*1000000000)
(69, 0.2284335353694217/250000000*1000000000)
(82, 0.23565244916628014/250000000*1000000000)
(89, 0.23892275383595796/250000000*1000000000)
(96, 0.23892275383595796/250000000*1000000000)
};
\addlegendentry{With hardware metrics}

\end{semilogyaxis}
\end{tikzpicture}    
    \caption{Comparison between software metrics-only sparsity search and the proposed hardware-aware sparsity search for ResNet-18. We use 96 iteration steps for both approaches.}
    \label{fig:co_dse}
\end{figure}
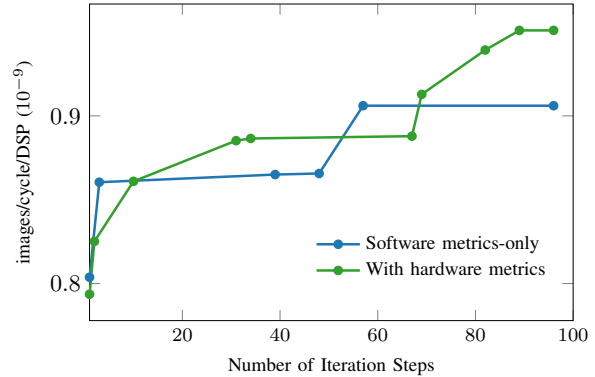
\label{subsec:multi}
Let's denote the per-layer weight and activation pruning threshold as $\tau_w$ and $\tau_a$. To identify the optimal values of them, we construct the following optimization problem for the network $L$:
\begin{itemize}
    \item $f_{acc}$: network accuracy, measured on validation data;
    \item $f_{spa}$: average sparsity of the network, including both weights and activations;
    \item $f_{thr}$: network throughput for a given pruned network searched by DSE;
    \item $f_{dsp}$: resource utilization of the accelerator, represented by the number of DSPs used, which is usually the bottleneck in dataflow accelerators.
\end{itemize}
The objective for the hardware-aware pruning is to maximize accuracy, sparsity, and throughput while minimizing resource utilization:
\begin{equation}
    \max_{\{\tau_w, \tau_a\}, l\in L} f_{acc}(L) + \lambda_1f_{spa}(L) + \lambda_2f_{thr}(L) - \lambda_3f_{dsp}(L)
    \label{equ:cost}
\end{equation}
$\lambda_1$, $\lambda_2$ and $\lambda_3$ are hyperparameters that normalize the values of these metrics, determined by heuristics. An efficient search algorithm for multi-objective search is required, and we use Tree-structured Parzen Estimator (TPE) \cite{bergstra2011algorithms}, a Bayesian optimization algorithm that uses a tree structure to model the probability density of the objective function and iteratively constructs this tree based on observed evaluations of the objective function.

The proposed hardware-aware pruning approach could achieve better computation efficiency (throughput per area) than the same pruning approach with software metrics only (i.e. accuracy and sparsity). Fig.~\ref{fig:co_dse} shows a comparison of these two approaches for ResNet-18. The proposed hardware-aware search shown as the green curve is guided by the objective function in Equation~\ref{equ:cost}. The software metrics-only search shown as the blue curve only uses accuracy and sparsity as the objective function. Both approaches ran 96 iteration steps, taking about 3 hours to complete. At the initial stage, the change in green curve in computation efficiency is slow because the objective function is complex and contains more hardware metrics. Then it reaches a better computation efficiency because the hardware metrics guided the search towards efficient hardware implementation. This is more helpful if more iteration steps are performed for large networks.
\begin{figure}
\centering
\pgfplotsset{compat=1.3}
\begin{tikzpicture}
\pgfplotsset{every x tick label/.append style={font=\footnotesize}}
\begin{axis}[
    width=80mm,
    ybar, height=55mm, ymode=log,
    legend style={draw=none},
    ymin=0, ymax=40,
    ylabel={\footnotesize images/cycle/DSP ($10^{-9}$)},
    legend style={font=\footnotesize},
    bar width=4,
    nodes near coords,
    log ticks with fixed point,
    ytick={0, 0.1, 0.2, 0.5, 1, 2},
    major x tick style = transparent,
    legend cell align={left},
    symbolic x coords={
ResNet-18, ResNet-50, MobileNetV2, MobileNetV3S, MobileNetV3L
    },
    xticklabels={ResNet-18, ResNet-50, MobileNetV2, MobileNetV3S, MobileNetV3L},
    xtick=data,
    xticklabel style={xshift=-15pt, yshift=5pt, rotate=20},
    nodes near coords,
    point meta=explicit symbolic,
    nodes near coords style={
            align=center,
            font=\footnotesize,
        },
    legend style={at={(0.05,0.95)},anchor=north west, draw=none, fill=none, font=\footnotesize},
    ]

\addplot[jcbluel, draw=jcbluel, fill=jcbluel] table
[x=model, y=base, col sep=space] {
model base ours times
ResNet-18 0.24 0.92 3.8$\times$
ResNet-50 0.08 0.42 5.3$\times$
MobileNetV2 1.28 3.42 2.7$\times$
MobileNetV3S  4.57 10.90 2.4$\times$
MobileNetV3L 1.15 1.76 1.5$\times$
};
\addlegendentry{Dense Dataflow}

\addplot[jcblue, draw=jcblue, fill=jcblue, 
] table
[x=model, y=ours, meta=times, col sep=space] {
model base ours times
ResNet-18 0.24 0.92 3.8$\times$
ResNet-50 0.08 0.42 5.3$\times$
MobileNetV2 1.28 3.42 2.7$\times$
MobileNetV3S  4.57 10.90 2.4$\times$
MobileNetV3L 1.15 1.76 1.5$\times$
};
\addlegendentry{Proposed Sparse Dataflow}

\end{axis}
\end{tikzpicture}
    \caption{By exploiting sparsity in both weights and activations, our approach achieves significant speedup compared to the dense implementation.}
    \label{fig:bar_chart}
\end{figure}
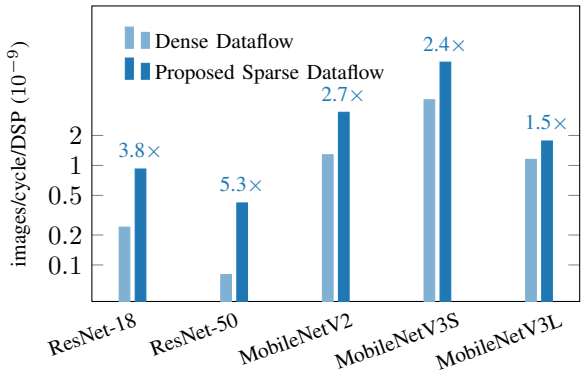

\section{Evaluation}
\label{sec:eva}

\begin{table*}[t]
    \centering
    \caption{A comparison with state-of-the-art sparse DNN-FPGA accelerators.}
    \label{tab:compare}
    \resizebox{\textwidth}{!}{%
    \begin{tabular}{lrrrrrrrrrrrrrrr}
\toprule
Models & \multicolumn{2}{c}{ResNet-18} & \multicolumn{3}{c}{ResNet-50} & \multicolumn{4}{c}{MobileNetV2} & \multicolumn{2}{c}{MobileNetV3S} & \multicolumn{2}{c}{MobileNetV3L} \\
\cmidrule(lr){2-3}
\cmidrule(lr){4-6}
\cmidrule(lr){7-10}
\cmidrule(lr){11-12}
\cmidrule(lr){13-14}
Works & \multicolumn{1}{c}{\cite{montgomerie2023pass}} & \multicolumn{1}{c}{Ours} & \multicolumn{1}{c}{\cite{liu2023efficient}} & \multicolumn{1}{c}{\cite{montgomerie2023pass}} & \multicolumn{1}{c}{Ours} & \multicolumn{1}{c}{\cite{liu2023efficient}} & \multicolumn{1}{c}{\cite{hall2020hpipe}} & \multicolumn{1}{c}{\cite{montgomerie2023pass}} &  \multicolumn{1}{c}{Ours} & Dense & \multicolumn{1}{c}{Ours} & Dense &  \multicolumn{1}{c}{Ours} \\ \midrule
Accuracy & 69.75 & 69.59 & N/A & 76.13 & 75.58 & 70.80 & 71.90 & 71.79 & 71.45 & 67.42 & 67.28 & 74.04 & 73.76 \\
Platform & U250 & U250 & 7V690T & U250 & U250 & 7V690T & Stratix10 & U250 & U250 & U250 & U250 & U250 & U250 \\
Freq (MHz) & 250 & 250 & 150 & 250 & 250 & 150 & 390 & 250 & 250 & 250 & 250 & 250 & 250 \\
DSPs & 10974 & 12234 & 2160 & 11952 & 7434 & 2160 & 5928 & 3596 & 5261 & 4282 & 1796 & 6577 & 4324 \\
kLUTs & 1659 & 1679 & 308 & 1721 & 1724 & 308 & 523 & 1552 & 1720 & 971 & 507 & 1535 & 1728 \\
BRAM18k & 4554 & 4817 & 1883 & 4262 & 4178 & 1883 & 4512 & 1774 & 1902 & 2278 & 1779 & 3706 & 5376 \\
images/s & 1904 & 2819 & 33 & 330 & 776 & 302 & 4539 & 1660 & 4495 & 4890 & 4895 & 1897 & 1898 \\
images/cycle/DSP ($10^{-9}$) & 0.69 & \textbf{\textcolor{jcgreen}{0.92}} & 0.10 & 0.11 & \textbf{\textcolor{jcgreen}{0.42}} & 0.93 & 1.96 & 1.84 & \textbf{\textcolor{jcgreen}{3.42}} & 4.57 & \textbf{\textcolor{jcgreen}{10.90}} & 1.15 & \textbf{\textcolor{jcgreen}{1.76}}\\ 
\bottomrule
\end{tabular}
}
\end{table*}

In terms of the experiment set-up, the FPGA device we used for the result measurements is AMD Xilinx Alveo U250, and the version of software employed is Vitis 2023.1. The clock frequency of our accelerators is 250MHz. We examine our approach on a set of mainstream DNN benchmarks including ResNet-18, ResNet-50, MobileNetV2 and MobileNetV3, taken from pyTorch torchvision and trained on the ImageNet dataset. All the networks quantized to 16-bit fixed point for both weights and activations. Our pruning algorithm is one-shot and post-training, without applying any fine-tuning to the networks.

The tool flow is presented in Fig.~\ref{fig:our_flow}. We implemented a fully-automated flow that translates DNN models to Torch FX graph. In an FX graph, the model is represented as a dataflow graph similar to the one shown on the left of Fig.~\ref{fig:sparse_engine}. Our pruning search engine extracts the layer-wise pruning space using the TPE algorithm. With the proposed hardware model presented in Section~\ref{subsec:thr}, the hardware cost is evaluated as well as the accuracy loss. This guides the next iteration step of the pruning search process. The final pruned model and its hardware parameters are sent to a hardware synthesis tool for hardware implementation. Our approach is general and can be extended to support other hardware synthesis tools. For this work, we used a DNN-FPGA synthesis tool named fpgaConvNet~\cite{venieris_fpgaconvnet_2018} for prototyping.

We compare our approach with other sparse DNN-FPGA accelerators from related works. Our evaluation encompasses the trade-off among accuracy, resource, throughput (images/s) and hardware efficiency (images/cycle/DSP). Fig.~\ref{fig:bar_chart} shows the improvements from the dense dataflow architecture to the proposed sparse dataflow architecture, and the detailed results are shown in Table~\ref{tab:compare}. We made the following observations: 
\begin{itemize}
    \item Compared with related work using sparse computation, the dataflow architectures, including HPIPE\cite{hall2020hpipe}, PASS \cite{montgomerie2023pass} and Ours, have significant performance improvements over the non-dataflow design \cite{liu2023efficient} because of its deeply pipelined computation. For example, when targeting ResNet-50 and MobileNetV2, the advantage in terms of throughput per DSP can be up to 4.2 $\times$ and 3.7$\times$, respectively.
    \item The advantage of sparse dataflow architecture is not always free. Compared to the non-dataflow sparse accelerator design \cite{liu2023efficient}, the dataflow approach requires more resources (upto 3$\times$ DSPs and 5$\times$ LUTs). This is because the hardware dataflow pipeline requires extra logic and buffers in exchange for pipeline performance and data parallelism scalability. 
    \item For MobileNetV3, we also include the dense results in the table. Compared with the dense results, our sparse implementations achieve the same throughput with a reduced number of DSP utilized. The throughput remains similar because the designs are either LUT or BRAM bounded.
    \item PASS \cite{montgomerie2023pass} is the work related most to our solution, and it optimizes hardware for the activation sparsity. However, they do not explore weight sparsity or consider to drive the pruning strategy with hardware information. Compared with them, our approach achieves improved efficiency with 1.3$\times$, 3.8$\times$ and 1.9$\times$ on ResNet-18, ResNet-50 and MobileNetV2. Meanwhile, our post-training accuracy degradation is less than 0.6 percentage points.
    \item The variance in the results depends on the sensitivity of the models to data sparsity. The accuracy loss may be reduced with the help of fine-tuning, at the price of training time. 
\end{itemize}

\section{Conclusion}
In this paper, we propose a novel approach to exploit software and hardware co-optimization for sparsity, targeting the dataflow DNN-FPGA accelerator. We implement a hardware-aware sparsity search that considers both software pruning metrics and hardware performance for systematic optimization. We also exploit both static weight sparsity and dynamic activation sparsity for dataflow accelerator design. Over a set of DNN models, we achieve 1.3$\times$ to 4.2$\times$ efficiency compared to existing sparse designs. In terms of future work, we will delve into the integration of a more diverse range of pruning algorithms to further enhance applicability. 

\bibliographystyle{IEEEtran}
\bibliography{bibliography}

\end{document}